# REGENERATION TESTS OF A ROOM TEMPERATURE MAGNETIC REFRIGERATOR AND HEAT PUMP

by G. V. Brown and S. S. Papell

Lewis Research Center

Cleveland, Ohio 44135

# REGENERATION TESTS OF A ROOM TEMPERATURE MAGNETIC REFRIGERATOR AND HEAT PUMP

G. V. Brown and S. S. Papell

NASA Lewis Research Center

Cleveland, Ohio 44135

## ABSTRACT

A magnetic heat pump apparatus consisting of a solid magnetic refrigerant, gadolinium, and a liquid regenerator column of ethanol and water has been tested. Utilizing a 7T field, it produced a maximum temperature span of 80 K, and in separate tests, a lowest temperature of 241 K and a highest temperature of 328 K. Thermocouples, placed at intervals along the regenerator tube, permitted measurement of the temperature distribution in the regenerator fluid. No attempt was made to extract refrigeration from the device, but analysis of the temperature distributions shows that 34 watts of refrigeration was produced.

The concepts that make magnetic refrigeration possible well above liquid helium temperature were advanced in an earlier paper.[1] They are: (a) employ a solid ferromagnetic refrigerant, (b) use a rare-earth-based ferromagnet, and (c) employ a regenerative thermodynamic cycle. This report summarizes the experimental results obtained from a test apparatus based on these concepts. The test device easily produced the ranges of temperature required in each of the following practical applications: heat pumping (for space heating), air conditioning, refrigeration and process refrigeration down to 241 K (-32° C).

A diagram of the test apparatus is shown in figure 1. The refrigerant was 0.9 kg of gadolinium (Gd), formed in 1 mm thick plates, assembled with 0.5 mm spacing between plates in a cylindrical housing of 304 stainless steel with 5 cm outside diameter and 0.5 mm wall. The refrigerant assembly was suspended motionless at the center of a water-cooled solenoid,[2] the field of which can be raised from zero to 7T in 5 seconds and reduced from 7T to less than 0.1T in 6 seconds. The regenerator fluid column, 102 cm long and composed of 50 percent ethanol and 50 percent water, was contained in an insulated glass tube. Copper-constantan thermocouples, soldered onto 1 cm × 10 cm copper strips to increase the thermal contact area, where cemented (epoxied) circumferentially onto the outside of the tube. The number 1 thermocouple was soldered onto a copper disc at the bottom of the liquid column. The regenerator assembly was connected to a reversible motor that could raise and lower the regenerator at any frequency up to 0.05 Hz. The motion could also be stopped at any point; in particular, it was stopped when the regenerator was at its lowest and highest positions.

The cycle was performed as follows: (1) with the regenerator at its lowest position (Gd at the top of the fluid column), the field was turned on. This caused the Gd temperature to rise and to transfer heat to the fluid in which it was immersed. (2) The regenerator was raised until the Gd was at the bottom of the fluid column. During the motion the fluid cooled the Gd nearly to the temperature of the fluid at the bottom. (3) The field was turned off. This cooled the Gd, which in turn, cooled the lower-most fluid. (4) The regenerator was lowered until the Gd was back at the top of the fluid column. During the motion the fluid warmed the Gd nearly to the temperature of the uppermost fluid. Successive cycles extracted heat from the lower part of the regenerator and deposited heat in the upper part, gradually increasing the temperature span between top and bottom.

The thermocouple records from one sequence of 50 cycles are shown in figure 2. The liquid column was initially slightly below room temperature. The first ten cycles required an average of 60 seconds each; the last ten averaged about 66 seconds each. After 7 cycles the fluid at the bottom reached 273 K, and the final lowest temperature was 256 K. The heat pumped from the lower part of the column, plus the work input into the cycle, was deposited in the upper part of the column, the top of which reached an ultimate temperature of 322 K. As can be seen from figure 2, a steady state was not reached in 50 cycles; a greater temperature span would have been achieved by continued operation. However, temperature limitations on the homopolar generator that supplied power to the magnet forced an end to the test. The use of a superconducting magnet, which would be required in a practical application of these concepts, would permit indefinite operation, but the field change would.probably have to be obtained by moving the Gd into and out of the magnet.

The greatest temperature span that has been achieved in a single test was 80 K, between 248 K and 328 K.

In another type of test, a heat exchanger formed from copper tubing was placed in the top 10 cm of the regenerator fluid, as shown in the insert of figure 1. Cooling water was run through the tubing to remove the heat of magnetization, which was deposited in the upper part of the regenerator fluid during each cycle. The temperature at the top of the column was thus maintained at a nearly constant temperature. A thermocouple record from one of these tests is presented in figure 3. The fluid was all between 253 K and 258 K before the heat exchanger was inserted. The water circulating through it raised the temperature at the top thermocouple to 293 K. By the end of the test, however, the temperature distribution in the fluid has developed so that the fluid at the top thermocouple was at 294 K, and heat was rejected into the heat exchanger coil. Again, because of the magnet power supply limitation, the test did not reach fully steady state conditions.

The lowest temperature yet obtained was 241 K in a test in which the fluid was all initially at 283 K and the heat exchanger temperature was also 283 K.

# REFRIGERATION RATES

No attempt was made to extract refrigeration from the device in these tests. The reason is that the performance of the magnetic refrigerator is dependent mainly on the effectiveness of regeneration, because the amount of heat that must be regenerated is typically an order of magnitude larger than the heat extracted from a load. Consequently, these tests were designed to explore the regenerator behavior. The rate of refrigeration produced in the lower part of the regenerator column can, however, be found directly from the temperature distributions presented in the figures. The results of the test shown in figure 2 indicate an initial refrigeration rate of 34 watts and a final rate of about 6 watts. In a later test the initial rate surpassed 60 watts of refrigeration in a pure water regenerator fluid at an initial temperature of 300 K.

The efficiency of the refrigeration process could in principle be derived from the data in figure 2, if the test device were perfectly insulated from its surroundings. However, the heat leak from the hot end to the surroundings and from the surroundings to the cold end was too large to permit the efficiency to be determined. Heat conduction down through the liquid column itself was only a small fraction of a watt, even when the highest temperature gradient existed.

The temperature span and refrigeration power produced in these tests are substantially greater than any previously reported for magnetic refrigeration in any temperature range. But markedly higher performance than reported here would be achieved if two limitations of the test device were removed. The first is an unwanted mixing of the regenerator fluid that is caused when the Gd assembly and housing pass through it. Proper regeneration depends upon the production and maintenance of a smooth temperature gradient in the fluid. If colder fluid mixes with hotter fluid because of turbulence in the wake of the Gd assembly, there are obviously irreversible losses of refrigeration and efficiency. In early tests the maximum temperature span achieved was often limited to only 15 K or 20 K because of the laminar jets of fluid which issued from between the Gd plates and caused serious mixing. Therefore, several layers of stainless steel screen wire were added at each end of the assembly of Gd plates in order to break up such large-scale jet motion into smaller scale turbulence that dissipates quickly. Also, the ends of the Gd plates are streamlined to improve the flow pattern. These measures brought substantial improvement.

The second factor limiting performance is the amount of surface area available for heat transfer between the Gd and the fluid. An engineering analysis of the effectiveness of regeneration (to be published elsewhere) shows that in an economically competitive refrigerator, much more heat transfer area is required. The Gd must be divided into sheet that is at most about 0.01 cm thick (or into some other form, such as screen wire, with as much surface area per unit volume of Gd). With a tenfold increase in surface area, from that of the present device, and with a decrease in the size of the fluid-flow passages between the plates to the order of 0.01 cm or less, the operating speed of the refrigerator could be increased to 1 Hz or more. Refrigeration output would rise almost in proportion to speed; hence the 0.9 kg of Gd could produce on the order of 3 kW of refrigeration. Even finer division of the Gd would give higher output. It is interesting to note that Stirling engines have been built in which the regenerator material is still finer, i.e., 200 mesh screen with 0.0038 cm wire diameter.[3]

It is not clear from the present results whether the performance of the test device is limited more by mixing or by the available heat transfer area.

## CONCLUSIONS

These tests have demonstrated the efficacy of regeneration in a magnetic refrigeration cycle. Furthermore, the usefulness of Gd has been demonstrated over a large temperature range that exceeds the requirements for all common refrigeration and heat pumping applications near room temperature. Even though the heat absorbing capability of the magnetic material falls off rather rapidly away from its Curie point,[4] a temperature of 241 K was reached by using Gd, which has a 293 K ferromagnetic Curie point.

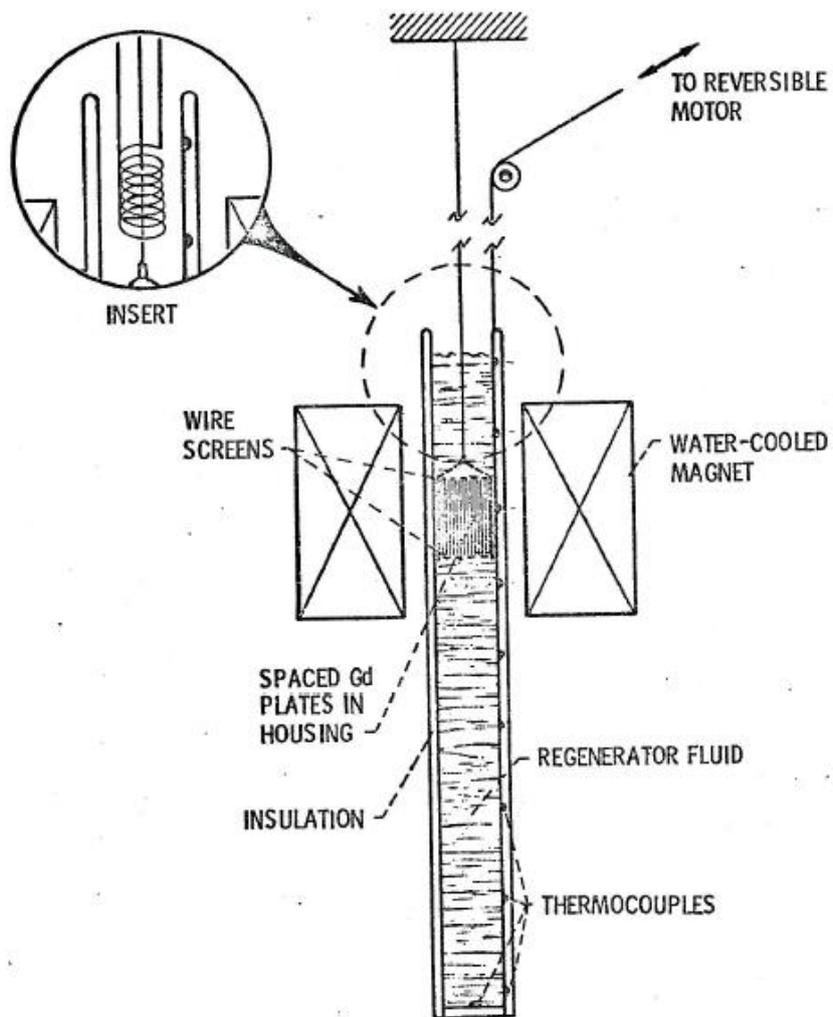

Figure 1. - Experimental apparatus. Magnet and Gd assembly are stationary. Field is turned on and off as required. The insert shows the position of the copper tubing heat exchanger when it is used.

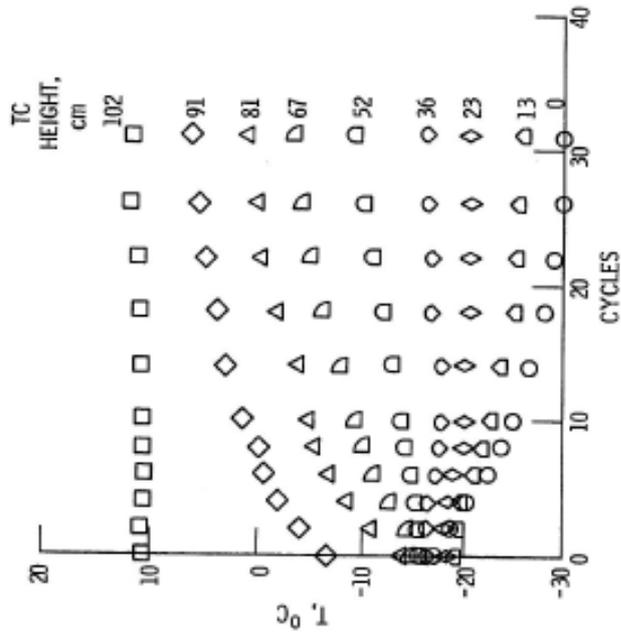

Figure 3. – Development of the temperature distribution in the regenerator fluid with the heat exchanger in the position shown in the insert of figure 1.

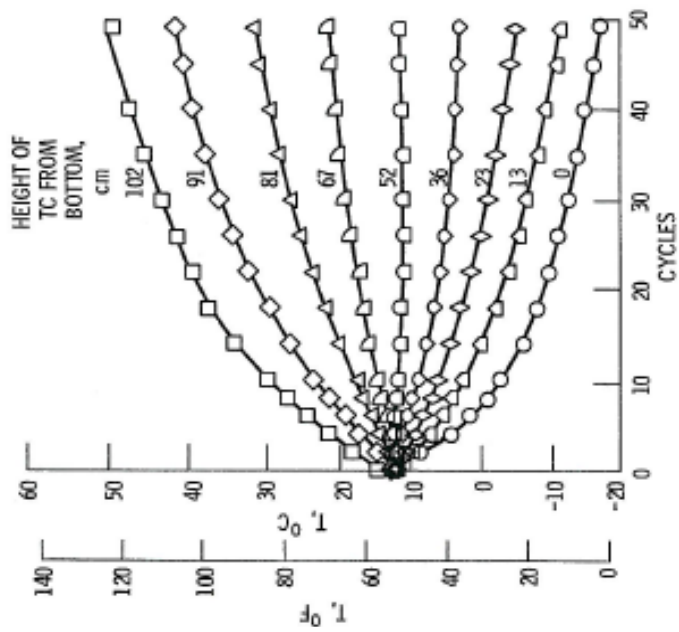

Figure 2. – Development of the temperature distribution in the regenerator fluid with no heat exchanger present.

LIST OF FIGURE CAPTIONS

Figure 1:   Experimental apparatus. Magnet and Gd assembly are stationary. Field is turned on and off as required. The insert shows the position of the copper tubing heat exchanger when it is used.

Figure 2:   Development of the temperature distribution in the regenerator fluid with no heat exchanger present.

Figure 3:   Development of the temperature distribution in the regenerator fluid with the heat exchanger in the position shown in the insert of figure 1.